# Black Phosphorus Radio-Frequency Transistors

Han Wang[1,*], Xiaomu Wang[2], Fengnian Xia[2,*], Luhao Wang[1], Hao Jiang[3], Qiangfei Xia[3], Matthew L. Chin[4], Madan Dubey[4], Shu-jen Han[5]

[1]Ming Hsieh Department of Electrical Engineering, University of Southern California, Los Angeles, CA 90089

[2]Department of Electrical Engineering, Yale University, New Haven, CT 06511

[3]Department of Electrical & Computer Engineering, University of Massachusetts, Amherst, MA 01003

[4] Sensors and Electron Devices Directorate, US Army Research Laboratory, Adelphi, MD 20723

[5]IBM T. J. Watson Research Center, Yorktown Heights, NY 10598

**Abstract.** Few-layer and thin film forms of layered black phosphorus (BP) have recently emerged as a promising material for applications in high performance nanoelectronics and infrared optoelectronics. Layered BP thin film offers a moderate bandgap of around 0.3 eV and high carrier mobility, leading to transistors with decent on-off ratio and high on-state current density. Here, we demonstrate the gigahertz frequency operation of black phosphorus field-effect transistors for the first time. The BP transistors demonstrated here show excellent current saturation with an on-off ratio exceeding $2\times10^3$. We achieved a current density in excess of 270 mA/mm and DC transconductance above 180 mS/mm for hole conduction. Using standard high frequency characterization techniques, we measured a short-circuit current-gain cut-off frequency $f_T$ of 12 GHz and a maximum oscillation frequency $f_{max}$ of 20 GHz in 300 nm channel length devices. BP devices may offer advantages over graphene transistors for high frequency electronics in terms of voltage and power gain due to the good current saturation properties arising from their finite bandgap, thus enabling the future ubiquitous transistor technology that can operate in the multi-GHz frequency range and beyond.

[*]Email: han.wang.4@usc.edu, fengnian.xia@yale.edu

# Introduction

Typical channel materials in thin film electronics include amorphous or polycrystalline silicon[1-3], organic compounds[4,5], and oxides[6,7]. These materials usually exhibit a sizable bandgap but compromised carrier mobility, making them undesirable for radio-frequency (RF) electronics. Early interests in using two-dimensional materials for RF thin film electronics focused mainly on graphene. Since the first demonstration of high frequency current gain in graphene field-effect transistors (FETs) in 2007[8], graphene has inspired great interests for RF applications due to its high mobility, high carrier velocity and long mean free path[9-14]. However, several years of intensive research has revealed that graphene transistors might suffer from a few fundamental limitations that will restrict its high frequency performance. The key shortcoming of graphene RF transistors is their lack of current saturation as a result of graphene's zero-bandgap nature, that can lead to reduced voltage and power gains[15,16], and to a lesser extent, the current gain[17]. As a result, most graphene RF FETs may have relatively high short-circuit current-gain cut-off frequency ($f_T$), but significantly lower maximum oscillation frequency ($f_{max}$), which benchmarks the power gain of the transistor. On the other hand, transistors based on transition metal dichalcogenides, such as molybdenum disulfide (MoS$_2$) and tungsten diselenide (WSe$_2$), only showed limited potential for high frequency applications because of their relatively low mobility[18,19]. Recently, layered black phosphorus (BP) thin film has emerged as a promising candidate for high performance thin film electronics due to its moderate bandgap of 0.3 eV and high carrier mobility. The recent experiment has shown hole mobility exceeding 650 cm$^2$ V$^{-1}$ s$^{-1}$ at room temperature and above 1,000 cm$^2$ V$^{-1}$ s$^{-1}$ at 120 K along the light effective mass (x) direction[20]. While in bulk black phosphorus, mobilities exceeding 1,000 cm$^2$ V$^{-1}$s$^{-1}$ at room temperature for both electrons and holes, and exceeding 50,000 cm$^2$ V$^{-1}$s$^{-1}$ at 30 K have been demonstrated[21]. In this work, we demonstrate for the first time the operation of black phosphorus FETs in the gigahertz frequency range. Future ubiquitous transistor technologies using this novel layered material with high mobility and highly desirable current saturation property may revolutionize the electronic systems for many civilian and defense applications.

## Results

### Device fabrication

Orthorhombic bulk black phosphorus is an elemental layered material with D$_{2h}$ point group symmetry as shown in Fig. 1a. It is the most stable allotrope of phosphorus and the electronic properties of bulk BP have been studied several decades ago[22-26]. Recently, BP in its few-layer and thin-film form has inspired renewed interests among physicists and engineers due to its promising potential for application in thin film electronics and infrared optoelectronics[20,27-29]. The layered nature of BP crystals allows single- and multi-layer atomic crystals to be obtained through mechanical exfoliation techniques. In our experiments, multi-layer BP was first exfoliated using the standard micromechanical cleavage method from a bulk BP crystal, then transferred onto 300 nm thick silicon dioxide thermally grown on highly resistive silicon (>10$^4$ Ω·cm). Atomic force microscope (AFM) can be used to determine the number of atomic layers in a BP flake, which has a layer-to-layer spacing of 0.53 nm[20,21]. The x-, y- and z-directions of the crystal lattice are indicated in Fig. 1a. In a series of recent publications, the mobility and the on-off current ratio of BP FETs have been studied with respect to the thickness and layer number in the BP channel[20,27,28]. In this work, BP thin films with thicknesses around 6-10 nm were selected for the best balanced mobility and on-off properties for RF applications. Transistors using thinner BP films exhibit an on-off current ratio as high as $2\times10^5$ but low mobility of below 100 cm$^2$ V$^{-1}$ s$^{-1}$ due to the scattering from the external environment. Utilization of excessively thick BP films can lead to higher mobility but a reduced on-off ratio and less pronounced current saturation. The inset of Fig. 1b shows the optical micrograph of a typical BP flake used for device fabrication. AFM measurements indicate a thickness of 8.5 nm in this flake (Fig. 1b). To enhance the channel mobility, crystal orientations of the flakes were first identified using either Raman spectroscopy or infrared spectroscopy techniques and the transistors were built along the light effective mass (x-) direction of the BP lattice. Fig. 1c shows the Raman spectrum of the BP flake with the excitation laser polarized along the x-direction. The characteristic peaks at 470, 440, and 365 cm$^{-1}$ correspond to $A_g^2$, $B_{2g}$, and $A_g^1$ modes, respectively[20,30]. The $A_g^2$ mode has higher intensity compared to $B_{2g}$, and $A_g^1$ modes with this particular excitation laser polarization. Fig. 1d shows the polarization-resolved infrared spectra of the flake and the x-direction can be clearly identified as the direction with the highest optical conductivity around the band edge[20,21]. The source and drain electrodes were formed with 1 nm Ti/20 nm Pd/ 30 nm Au metal stack, which favors p-type carrier injection into the channel owing to the large work-function (~5.22 eV) of Pd. The gate dielectric was made with 21 nm of HfO$_2$ deposited by atomic layer deposition (ALD) technique at a temperature of 150 °C. The typical dielectric constant is around 13, determined by ellipsometric measurement on a control sample. Finally, the gate electrode was defined by electron-beam lithography to form transistors with sub-micrometer channel lengths. The fabrication process is described in more detail in Supporting Information. Fig. 2b shows the optical micrograph of the full layout of the device. The standard ground-signal-ground (GSG) pads were fabricated to realize signal transition from microwave coax cables to on-chip coplanar waveguide electrodes.

**DC Characterization**

Fig. 2a shows the schematic of the transistor structure. Fig. 2c and 2d display the DC characteristics of a top-gated BP FET with a 300 nm channel length ($L_G$). The device was built along the x-direction of the 8.5 nm thick flake shown in Fig. 1. Fig. 2c shows the measured drain current ($I_{DS}$) as a function of drain-source bias voltage ($V_{DS}$) at gate bias ($V_{GS}$) from 0 V to -2 V in steps of -0.5 V. The current saturation in the BP transistor is clearly visible and significantly improved from that in most graphene FETs due to the finite bandgap in BP thin film. This is typical for a BP device with ~8.5 nm channel thickness and agrees well with previous demonstrations[20]. Good current saturation characteristics will lead to low output conductance defined as $g_0 = \frac{dI_{DS}}{dV_{DS}}\big|_{V_{GS}=\text{constant}}$, which is the differential drain current change with respect to the variation in drain voltage bias. Low output conductance is critical for improving voltage and power gains, and to a lesser extent, the current gain in RF transistors. As reported in our previous work[20], typical Hall mobility along x-direction of BP flakes with thickness around 8 nm is above 400 cm$^2$ V$^{-1}$s$^{-1}$ at room temperature. Fig. 2d shows $I_{DS}$ as a function of $V_{GS}$ at $V_{DS}$=-2 V, where on-off current ratio over $2\times10^3$ is achieved. The as-fabricated device shows p-type conduction with threshold voltage around -0.7 V. A key factor influencing the high-frequency small signal response of a transistor is its transconductance ($g_m$), defined as the first derivative of the transfer characteristics, i.e. $g_m = \frac{dI_{DS}}{dV_{GS}}\Big|_{V_{DS}=constant}$. The inset of Fig. 2d shows the measured $g_m$ as a function of the gate voltage at $V_{DS}$=-2 V. The peak value of this extrinsic $g_m$ exceeds 180 mS/mm at $V_G$=-1.75 V while the peak on-state current density measured exceeds 270 mA/mm at $V_{DS}$=-2 V and $V_{GS}$=-2.5 V. Fig. 2d also shows $I_{DS}$ as a function of top-gate voltage $V_{GS}$ at a drain bias of $V_{DS}$=-2 V with the current plotted in logarithmic scale. The on-off current ratio of the device exceeds $2\times10^3$. Hence, the BP transistor demonstrated here shows significant advantages over graphene transistors in terms of its current saturation properties and on-off current ratio. Typical monolayer graphene devices have on-off current ratio less than 10 at room temperature. In bilayer graphene devices, the on-off current ratio only reaches around 100 even with bandgap opening induced by a strong external electrical field[31].

**RF Characterization**

To characterize the high frequency performance of BP RF transistors, we used the standard S-parameter measurement with on-chip probing utilizing GSG probes and Agilent N5230 vector network analyzer up to 50 GHz. Key figures of merit for microwave transistors can then be obtained for the BP devices. The network analyzer and the entire testing fixture were first calibrated using standard open, short, and load calibrations. Standard open and short structures were then used to de-embed the signals of the parallel and series parasitics associated with the measurement pads and connections[15,32-34]. The measurement procedure used in this work followed strictly the standard calibration and de-embedding processes widely accepted in the semiconductor industry, where the calibration step moves the reference plane to the tips of the GSG probes and the de-embedding step gives access to the performance of the active device region.

Fig. 3a and 3b plot the short-circuit current gain ($h_{21}$), the unilateral power gain (U) and maximum stable gain (MSG)/maximum available gain (MAG) extracted from S-parameters before and after de-embedding, respectively, measured at $V_{DS}$=-2.0 V and $V_{GS}$=-1.7 V for the device with $L_G$=300 nm. The plot of $|h_{21}|^2$ follows the characteristic $1/f$ relation with respect to frequency at a 20 dB/dec slope[16]. Fig. 3a and 3b show that the 300 nm channel length device has a $f_T$ of 7.8 GHz before de-embedding and 12 GHz after de-embedding, as extracted from the frequencies at which $|h_{21}|$ reaches unity. Gummel's method provides another way of extracting the cut-off frequency[35] where $f_T$ is extracted from the reciprocal of the initial slope in the imaginary part of $1/h_{21}$ vs. frequency plot. The values of $f_T$ obtained by both $1/f$ extraction and Gummel's method match closely for both 300 nm (Fig. 3c and 3d) and 1 µm (see Fig. S1 in the Supporting Information) channel length devices before and after de-embedding.

While $f_T$ is an important figure of merit related to the intrinsic speed of the BP transistor, another key figure of merit for analog application is the maximum oscillation frequency. It is the highest possible operating frequency at which a transistor can still amplify power. $f_{max}$ can be extracted from U or MSG/MAG of the device[16,36]. The unilateral power gain, U, also known more generally as Mason's U invariant, is a key parameter for any general two-port network. It carries great significance as an invariant parameter of the system under linear, lossless and reciprocal transformations. In Mason's classic work[37,38], the rich physical meaning of U was interpreted in three different ways as a maximum power gain, as a device activity measure, and as an invariant under a class of bilinear Möbius transformations. In transistor characterizations, U is the power gain under the condition of (1) *Unilateralization*, and (2) *Conjugate matched load for maximum power transfer*. In Fig. 3a and 3b, the plots of U follow a 20 dB/dec slope, and both U and MSG/MAG plots give similar $f_{max}$ of 12 GHz before de-embedding and 20 GHz after de-embedding, respectively.

The 300 nm channel length device has an extrinsic $f_T \cdot L_G$ product of 3.6 GHz μm. Using a slightly different design of the open pattern where the gate electrode in the "open" structure extends into the spacing between the source and drain electrodes[15], we can eliminate most of the gate-source and gate-drain parasitic capacitances $C_{gs}$ and $C_{gd}$. This will allow the extraction of a new $f_T$ value that reflects the more intrinsic property of the BP channel, and an intrinsic $f_T$ value ($f_{T,int}$) close to 51 GHz is obtained for the same device, which corresponds to the average saturation velocity in the channel approximately equal to $v_{sat}=2\pi f_T \cdot L_G \sim 9.6 \times 10^6$ cm/s. However, we would like to emphasize that $f_{T,int}$ only represents the upper limit of the possible frequency spectrum for this transistor. In any practical applications, the $C_{gs}$ and $C_{gd}$ of the device always significantly affect the device performance. As a result, we report $f_T$=12 GHz and $f_{max}$=20 GHz in Fig. 3 as the practically operable cut-off frequencies of the active device region[15,33], which are extracted based on standard characterization techniques commonly used for the characterization in silicon and III-V high frequency transistors. The intrinsic cut-off frequency of 51 GHz is extracted only as a way to approximately estimate the saturation velocity and it may not be appropriate for technology benchmarking. The RF characteristics for a device with 1 μm channel length are also reported in the Supporting

Information. The 1 μm channel length device has peak $f_T$=2.8 GHz and $f_{max}$=5.1 GHz before de-embedding and $f_T$=3.3 GHz and $f_{max}$=5.6 GHz after de-embedding (see Fig. S1 in the Supporting Information).

The intrinsic limit of the cut-off frequency can be estimated using $f_T = \frac{g_m}{2\pi C_{gs}}$. Assuming the same $g_m$ and gate dielectrics property from the measured data, an $f_T$ above 100 GHz may be reached if the channel length reduces to 30 nm. On the other hand, $f_{max}$ depends on both the current and voltage gains. It is related to $f_T$, following $f_{max} = \frac{f_T}{2}\sqrt{\frac{r_0}{R_g+R_i}}$. We can see that a high $f_T$ will enhance $f_{max}$ and good current saturation, i.e. high output resistance ($r_0$), low gate resistance ($R_g$) and low input resistance ($R_i$) are critical for improving $f_{max}$ of the device.

Fig. 4 shows the magnitude of the small-signal open-circuit voltage gain $|z_{21}/z_{11}|$ as a function of frequency before and after de-embedding. An open-circuit voltage gain refers to the voltage gain subject to infinitely high load impedance. Here, the voltage gain is obtained from the ratio between the open circuit forward transfer impedance $z_{21}$ and the open circuit input impedance $z_{11}$. Since $z_{11} = \left.\frac{v_1}{i_1}\right|_{i_2=0}$ and $z_{21} = \left.\frac{v_2}{i_1}\right|_{i_2=0}$, the ratio $z_{21}/z_{11}$ is equal to $\left.\frac{v_2}{v_1}\right|_{i_2=0}$, i.e. the voltage gain with the output port in open condition. $v_1$ and $v_2$ refer to the voltages at the input and output ports, respectively. In transistors, $v_1$ is the small-signal input voltage at the gate and the $v_2$ is the small-signal output voltage at the drain. Based on the definition of $g_m$ and $g_0$ discussed earlier, we can see that the voltage gain $|z_{21}/z_{11}|$ is closely related to the ratio $g_m/g_0$. Devices with good current saturation and high transcondutance are hence expected to have high voltage gains. The BP transistors show good voltage gain characteristics. As shown in Fig. 4, the before-de-embedding voltage gain stays above unity (0 dB) up to 13 GHz. The after-de-embedding voltage gain is close to 20 dB at 2 GHz and stays above unity (0 dB) for the entire frequency range measured up to 50 GHz. For a longer channel length (1 μm) device (see Fig. S2 in the Supporting Information), the before-de-embedding voltage gain is above unity (0 dB) up to 10 GHz. The after-de-embedding voltage gain is above 15 dB at 1 GHz and stays above unity (0 dB) up to 30 GHz.

## Summary

In this work, we investigated the high frequency characteristics of black phosphorus field effect transistors, whose channels were fabricated along the light effective mass (x-) direction. The device operates well into the GHz frequency range of the radio frequency spectrum. We carried out standard S-parameter measurements to characterize the high frequency response of these top-gated BP transistors. The short-circuit current gain, maximum stable gain/maximum available gain, unilateral power gain and voltage gain of the devices were carefully extracted. The short-circuit current gain of BP transistors shows the 20 dB/dec $1/f$ frequency dependence at high frequency. We measured a peak short-circuit current gain cutoff frequency $f_T$ of 12 GHz and maximum oscillation frequency $f_{max}$ of 20 GHz for a 300 nm channel length BP transistor, demonstrating the GHz operation of BP devices for the first time. These results clearly reveal the potentials of BP transistors to function as power and voltage amplifiers in multi-GHz frequency analogue and digital electronics demanded by many emerging civilian and military applications.

**Competing financial interests**

Authors declare no competing financial interests.

**Supporting Information**

Information on the device fabrication process and characterization method, additional high frequency characterization data are included in the Supporting Information. This material is available free of charge via the Internet at http://pubs.acs.org.

**Additional information**

Correspondence and requests for materials should be addressed to H.W. (han.wang.4@usc.edu) and F. X. (fengnian.xia@yale.edu)

**Figure Captions**

**Figure 1. Characterization of the black phosphorus (BP) thin film.**

(a) Layered crystal structure of black phosphorus. The spacing between adjacent layers is 5.3 Å. (b) atomic force microscope (AFM) data showing the thickness of a BP flake. The inset shows the optical micrograph of the BP flake with thickness around 8.5 nm. (c) Raman spectrum of the BP flake along x-direction. (d) Polarization-resolved infrared spectra of a BP flake.

**Figure 2. DC characteristics of BP transistors.**

(a) Schematic of the BP transistor device structure. (b) Optical micrograph of the fabricated device. (c) Output characteristics of the BP transistor. $L_G$= 300 nm. (d) Transfer characteristics of the same BP transistor plotted in both linear and logarithmic scale. $V_{DS}$=-2 V. The device has an on-off current ratio exceeding $2\times10^3$. The inset shows the transconductance $g_m$ of the device.

**Figure 3. Current and power gain in BP transistors at GHz frequency.**

(a) and (b) the short-circuit current gain $h_{21}$, the maximum stable gain and maximum available gain MSG/MAG and the unilateral power gain U of the 300 nm channel length device before and after de-embedding, respectively. The device has $f_T$=7.8 GHz, $f_{max}$=12 GHz before de-embedding, and $f_T$=12 GHz, $f_{max}$=20 GHz after de-embedding. (c) and (d) the imaginary part of $1/h_{21}$ as a function of frequency before and after de-embedding, respectively. Based on Gummel's method, the initial slope of the curve is equal to $1/f_T$.

**Figure 4. Open-circuit voltage gain in BP transistors at GHz frequency**

The open-circuit voltage gain ($z_{21}/z_{11}$) before and after de-embedding is shown as a function of the frequency. After de-embedding, the voltage gain stays close to 20 dB up to 2 GHz and is above unity (0 dB) in the entire measurement range up to 50 GHz. The grey dashed line is a guide to the eyes.

# Figure 1. Characterization of the black phosphorus (BP) thin film

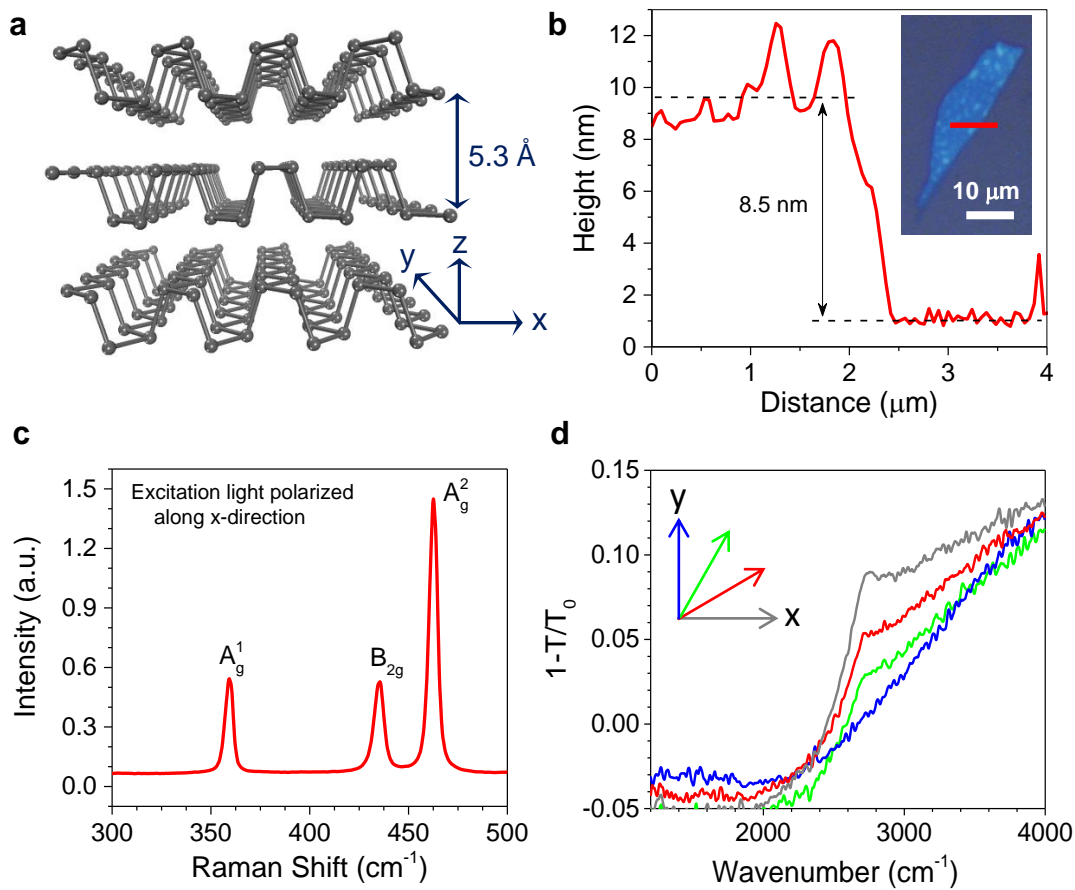

# Figure 2. DC characteristics of BP transistors

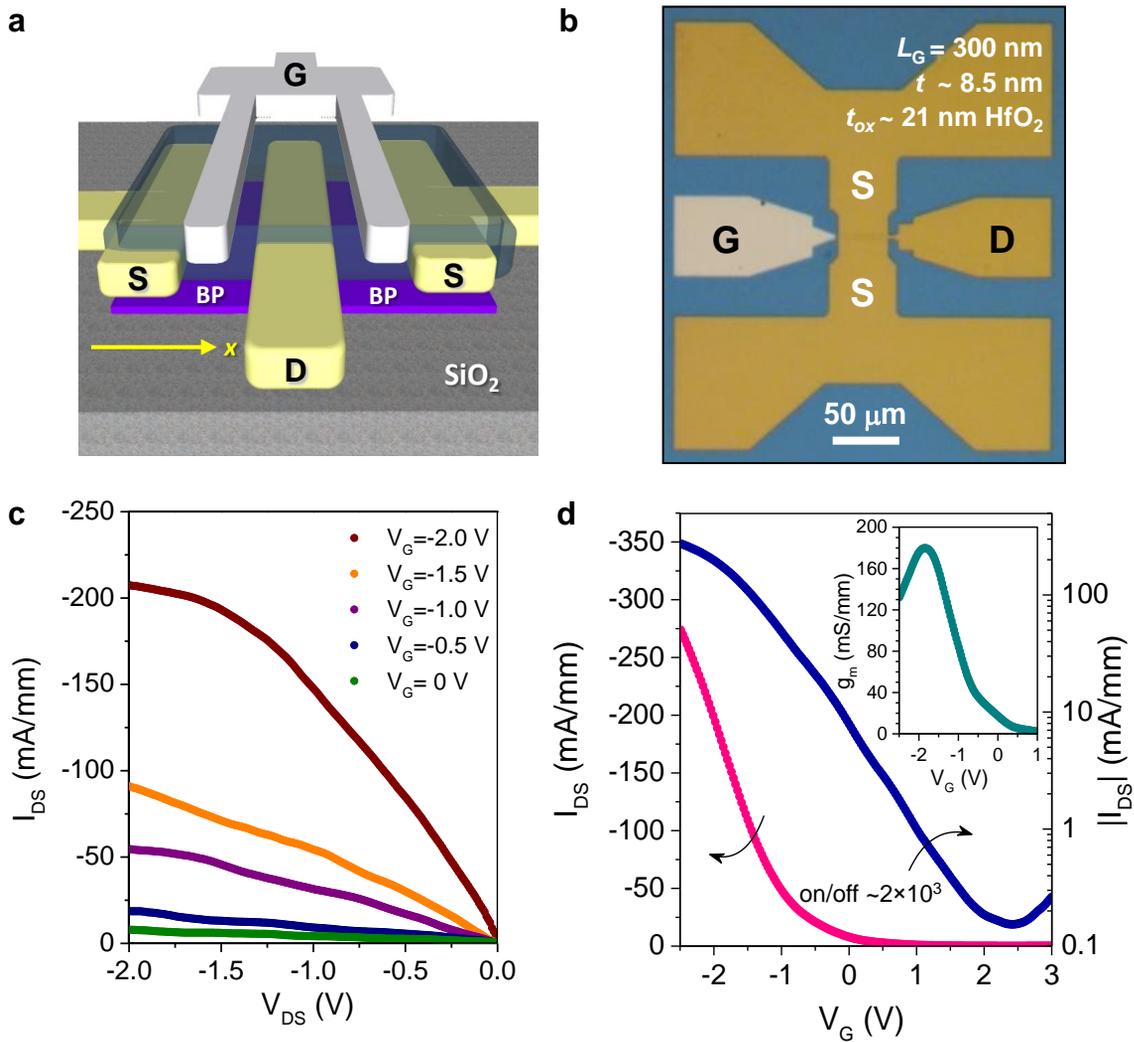

# Figure 3. Current and power gain in BP transistors at GHz frequency

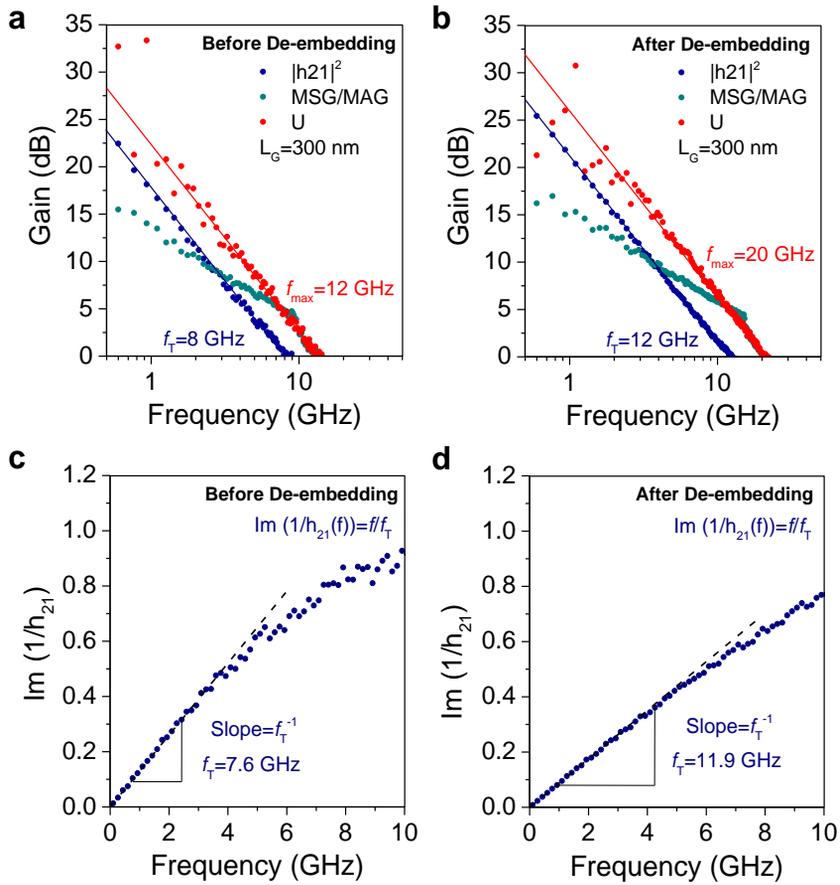

# Figure 4. Open-circuit voltage gain in BP transistors at GHz frequency

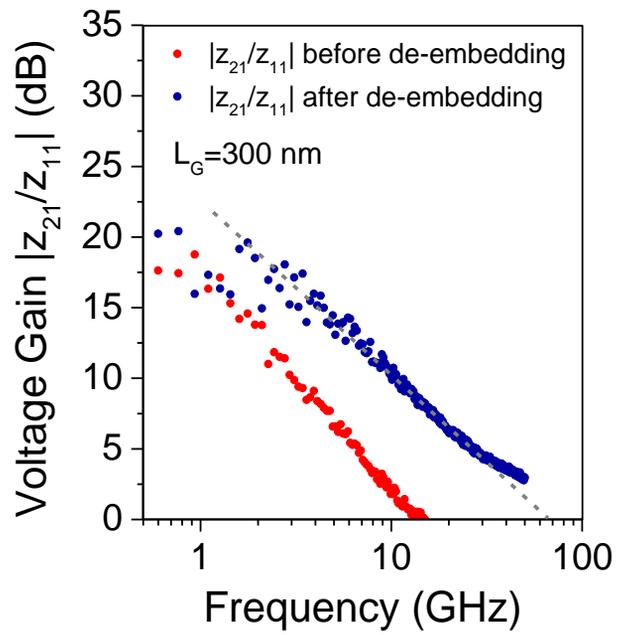

Supporting Information for

# Black Phosphorus Radio-Frequency Transistors

Han Wang[1,*], Xiaomu Wang[2], Fengnian Xia[2,*], Luhao Wang[1], Hao Jiang[3], Qiangfei Xia[3], Matthew L. Chin[4], Madan Dubey[4], Shu-jen Han[5]

[1]Ming Hsieh Department of Electrical Engineering, University of Southern California, Los Angeles, CA 90089

[2]Department of Electrical Engineering, Yale University, New Haven, CT 06511

[3]Department of Electrical & Computer Engineering, University of Massachusetts, Amherst, MA 01003

[4] Sensors and Electron Devices Directorate, US Army Research Laboratory, Adelphi, MD 20723

[5]IBM T. J. Watson Research Center, Yorktown Heights, NY 10598

[*]Email: han.wang.4@usc.edu, fengnian.xia@yale.edu

## Methods

**Top-gated transistor fabrication.** The fabrication of our devices starts with the exfoliation of BP thin films from bulk BP crystals onto 300 nm $SiO_2$ on a Si substrate, which has pre-patterned alignment grids, using the micro-mechanical cleavage technique. The thickness of the $SiO_2$ was selected to provide the optimal optical contrast for locating BP flakes relative to the alignment grids. The thickness of BP layers was measured by atomic force microscopy (AFM). The next step was to pattern the resist layer for metallization using a Vistec 100 kV electron-beam lithography system based on poly (methyl methacrylate) (950k MW PMMA). PMMA A3 was spun on wafer at a speed of 3000 rpm for 1 minute and was then bakes at 175 degrees for 3 minutes. The dose for exposure is 1100 µC $cm^{-2}$. Development was performed in 1:3 MIBK: IPA (Methyl isobutyl ketone: Isopropanol) for 90 s. We then evaporated 1 nm Ti/20 nm Pd/30 nm Au followed by lift-off in acetone to form the contacts. The $HfO_2$ gate dielectric is deposited using atomic layer deposition (ALD) at 150 °C. The gate electrode is also patterned using Vistec 100 kV electron-beam lithography system.

**AFM.** Atomic force microscopy (AFM) for identifying the thin film thickness was performed on a Digital Instruments/Veeco Dimension 3000 system.

**IR spectroscopy.** Bruker Optics Fourier Transfer Infrared spectrometer (Vertex 70) integrated with a Hyperion 2000 microscope system was used to measure the infrared spectroscopy of the BP flakes in the 800 $cm^{-1}$ to 4000 $cm^{-1}$ range. The linear polarization of the incident light was achieved using an infrared polarizer.

**Electrical characterization.** DC electrical characterizations were performed using an Agilent B1500 semiconductor parameter analyzer and a Lakeshore cryogenic probe station with micromanipulation probes. The high frequency characterizations were performed using an Agilent N5230A Vector Network Analyzer.

**High frequency characterization of BP transistor with 1 μm channel length**

The short-circuit current and power gain of a BP transistor with 1 μm channel length are shown in Fig. S1. The open-circuit voltage gain $z_{21}/z_{11}$ both before and after de-embedding are shown in Fig. S2. The device has $f_T$=2.8 GHz, $f_{max}$=5.1 GHz before de-embedding, and $f_T$=3.3 GHz, $f_{max}$=5.6 GHz after de-embedding. After de-embedding, the voltage gain stays above 15 dB up around 1 GHz and is above unity (0 dB) up to 30 GHz.

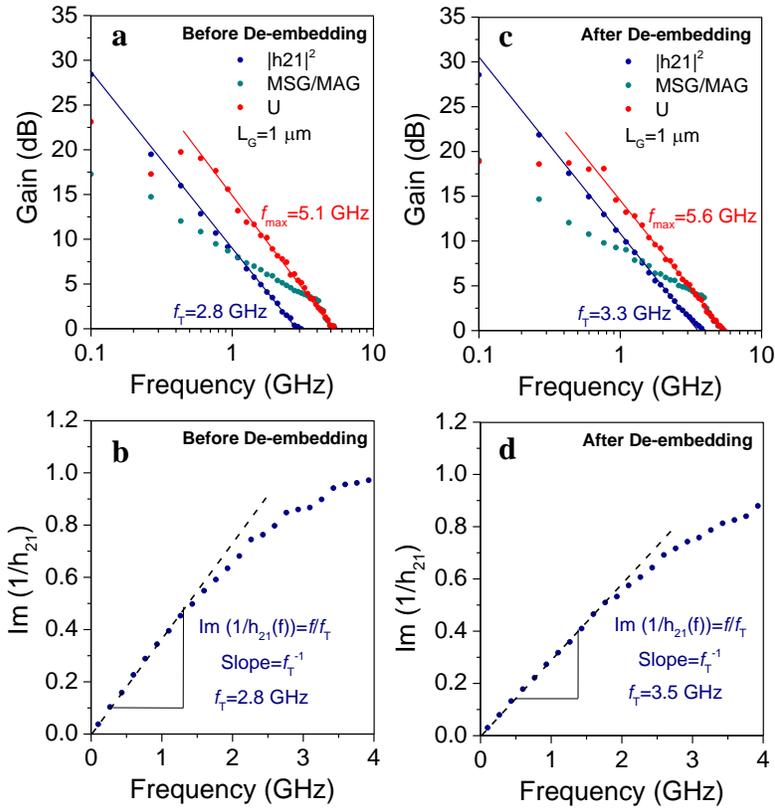

Fig. S1 (a) and (b) the short-circuit current gain $h_{21}$, the maximum stable gain and maximum available gain MSG/MAG and the unilateral power gain U of the device with 1 μm channel length before and after de-embedding, respectively. The device has $f_T$=2.8 GHz, $f_{max}$=5.1 GHz before de-embedding, and $f_T$=3.3 GHz, $f_{max}$=5.6 GHz after de-embedding. (c) and (d) the imaginary part of $1/h_{21}$ as a function of frequency before and after de-embedding, respectively. Based on Gummel's method, the initial slope of the curve is equal to $1/f_T$.

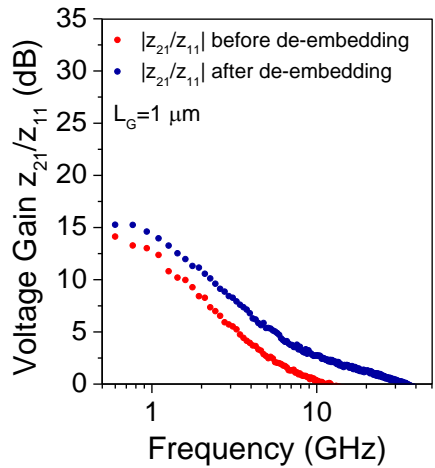

Fig. S2 the open-circuit voltage gain ($z_{21}/z_{11}$) before and after de-embedding is shown as a function of the frequency. After de-embedding, the voltage gain stays above 15 dB up to around 1 GHz and is above unity (0 dB) up to 30 GHz.